# Adaptive Bernstein Copulas and Risk Management

(corrected version)

Dietmar Pfeifer[1] and Olena Ragulina[2]

Carl von Ossietzky Universität Oldenburg, Germany[1] and
Taras Shevchenko National University of Kyiv, Ukraine[2]

**March 3, 2021**

**Abstract:** We present a constructive approach to Bernstein copulas with an admissible discrete skeleton in arbitrary dimensions when the underlying marginal grid sizes are smaller than the number of observations. This prevents an overfitting of the estimated dependence model and reduces the simulation effort for Bernstein copulas a lot. In a case study, we compare different approaches of Bernstein and Gaussian copulas w.r.t. the estimation of risk measures in risk management.

## 1. Introduction

Since the pioneering paper by Serge Bernstein in 1912 [3] Bernstein polynomials have been an indispensable tool in calculus and approximation theory (see e.g. [14]). Bernstein copulas, which can be considered as Bernstein polynomials for empirical and other copula functions, have a long tradition in non-parametric modelling of dependence structures in arbitrary dimensions, in particular with applications in risk management, and have come into a deeper focus in the recent years. There is an extensive list of research papers on this topic, in particular [2], [5], [6], [9], [10], [13], [16], [17], [22], [23], [24] and [25]. The monographs [8] and [11] have, in particular, devoted separate chapters to the topic of Bernstein copulas.

A very important aspect of Bernstein copulas lies in Monte Carlo simulation techniques of dependence structures, in particular in higher dimensions. The structure of such procedures ranges from very complex (see e.g. [17]) to extremely simple (see e.g. [6]) such that Monte Carlo simulations could e.g. be performed easily with ordinary spreadsheets, in particular in applications concerning quantitative risk management.

From a statistical point of view, the problem of a potential overfitting of the true underlying dependence structure with Bernstein polynomials emerges naturally. Clearly the Bernstein copula density becomes more wiggly the more empirical observations are used in the analysis. In comparison with classical parametric dependence models such as elliptically contoured or Archimedean copulas, this is probably a non-desirable property. This problem has in particular been tackled seemingly first in [17] by approximating the underlying discrete skeleton for the Bernstein copula by a least-squares approach and recently in the Ph.D. Thesis [1] where cluster analytic methods were used.

In the present paper, we propose a simple but yet effective approach to reduce the complexity of Bernstein copulas in a two-step approach, namely first an augmentation step in combination with a second reduction step. The reduction step is also discussed in [23], however without a possible application to a general complexity reduction of Bernstein copulas.

## 2. Some important facts about multivariate Bernstein polynomials

Let $f$ be an arbitrary bounded real-valued function on the unit cube $\mathbf{C}_d := [0,1]^d$ with dimension $d \in \mathbb{N}$. Let further $n_1, \cdots, n_d$ be integers. The corresponding multivariate Bernstein polynomial is defined by

$$B_{\mathbf{n}} f(x_1, \cdots, x_d) := \sum_{i_d=0}^{n_d} \cdots \sum_{i_1=0}^{n_1} f\left(\frac{i_1}{n_1}, \cdots, \frac{i_d}{n_d}\right) \prod_{j=1}^{d} \binom{n_j}{i_j} x_j^{i_j} (1-x_j)^{n_j - i_j}, \quad \mathbf{x} = (x_1, \cdots, x_d) \in \mathbf{C}_d \tag{1}$$

with $\mathbf{n} = (n_1, \cdots, n_d)$ (see e.g. [14], p. 51). It is known that for $\min(n_1, \cdots, n_d) \to \infty$ multivariate Bernstein polynomials converge to $f$ at any point of continuity and approximate $f$ uniformly if $f$ is continuous on $\mathbf{C}_d$.

Another important property of multivariate Bernstein polynomials that is perhaps less known in the mathematical community is the fact that the multivariate Bernstein polynomial density given by

$$b_{\mathbf{n}} f(x_1, \cdots, x_d) := \frac{\partial^d}{\partial x_1 \cdots \partial x_d} B_{\mathbf{n}} f(x_1, \cdots, x_d), \quad \mathbf{x} \in \mathbf{C}_d \tag{2}$$

can be written as a linear combination of statistical product beta densities. For this purpose, consider univariate beta densities

$$f_{\text{beta}}(x; \alpha, \beta) := \frac{x^{\alpha-1}(1-x)^{\beta-1}}{\mathrm{B}(\alpha, \beta)} \quad \text{for } 0 < x < 1, \ \alpha, \beta > 0 \tag{3}$$

where $\mathrm{B}(\alpha, \beta)$ denotes the Euler Beta-function, i.e. $\mathrm{B}(\alpha, \beta) = \frac{\Gamma(\alpha) \cdot \Gamma(\beta)}{\Gamma(\alpha + \beta)}$. We need a further definition to proceed.

**Definition 1.** Let $g$ be a real-valued bounded function on $\mathbb{R}^d$. We call

$$\Delta g_{\mathbf{a}}^{\mathbf{b}} := \sum_{(\varepsilon_1, \cdots, \varepsilon_d) \in \{0,1\}^d} (-1)^{\sum_{i=1}^{d} \varepsilon_i} g(\varepsilon_1 a_1 + (1-\varepsilon_1) b_1, \cdots, \varepsilon_d a_d + (1-\varepsilon_d) b_d) \geq 0 \tag{4}$$

the $\Delta$- difference of $g$ over the interval $(\mathbf{a}, \mathbf{b}] := \left(\underset{i=1}{\overset{d}{\times}} (a_i, b_i]\right)$ with $\mathbf{a} = (a_1, \cdots, a_d)$, $\mathbf{b} = (b_1, \cdots, b_d) \in \mathbb{R}^d$ and $a_i < b_i$, $1 \leq i \leq d$. (We adopt here a notation similar as in [15], Definition 2.1, which is slightly different from the notation in [7], Definition 1.2.10.)

**Proposition 1.** With the above notation, the Bernstein polynomial density $b_{\mathbf{n}} f$ can be represented as

$$b_{\mathbf{n}} f(x_1, \cdots, x_d) = \sum_{i_d=0}^{n_d-1} \cdots \sum_{i_1=0}^{n_1-1} \Delta f_{\mathbf{a}_{\mathbf{i}}}^{\mathbf{b}_{\mathbf{i}}} \prod_{j=1}^{d} f_{\text{beta}}(x_j; i_j + 1, n_j - i_j) \tag{5}$$

with $\mathbf{a}_{\mathbf{i}} := \left(\frac{i_1}{n_1}, \cdots, \frac{i_d}{n_d}\right)$ and $\mathbf{b}_{\mathbf{i}} := \left(\frac{i_1+1}{n_1}, \cdots, \frac{i_d+1}{n_d}\right)$.

**Proof.** This follows immediately from the arguments in the proof of Theorem 2.2. in [6]; compare also the line of proofs in [4]. ●

**Example 1.** We consider the polynomial $f(x,y) := 2x(1-y)^3 - 3(1-x)^3 y^4$, $0 \leq x, y \leq 1$ with $n_1 = 2$, $n_2 = 3$. In this case, the two-dimensional Bernstein polynomial $B_n f$ differs from $f$ due to smaller polynomial degrees. We have

$$B_n f(x,y) = 2x - \frac{y}{9} - \frac{145}{36} xy - \frac{14}{9} y^2 - \frac{4}{3} y^3 + \frac{97}{18} xy^2 - \frac{x^2 y}{12} + \frac{17}{9} xy^3 - \frac{7}{6} x^2 y^2 - x^2 y^3 \tag{6}$$

with

$$b_n f(x,y) = -\frac{145}{36} - \frac{x}{6} + \frac{97}{9} y + \frac{17}{3} y^2 - \frac{14}{3} xy - 6xy^2. \tag{7}$$

Note that here

$$\Delta f_{a_i}^{b_i} = f\left(\frac{i_1+1}{n_1}, \frac{i_2+1}{n_2}\right) + f\left(\frac{i_1}{n_1}, \frac{i_2}{n_2}\right) - f\left(\frac{i_1}{n_1}, \frac{i_2+1}{n_2}\right) - f\left(\frac{i_1+1}{n_1}, \frac{i_2}{n_2}\right) \tag{8}$$

or, in tabular form,

| $i_1$ | 0 | 1 | 0 | 1 | 0 | 1 |
|---|---|---|---|---|---|---|
| $i_2$ | 0 | 0 | 1 | 1 | 2 | 2 |
| $\Delta f_{a_i}^{b_i}$ | $-\frac{145}{216}$ | $-\frac{151}{216}$ | $\frac{49}{216}$ | $-\frac{41}{216}$ | $\frac{149}{72}$ | $\frac{19}{72}$ |

Tab. 1

After a little computation it is easy to see that indeed here

$$b_n f(x,y) = \sum_{i_2=0}^{2} \sum_{i_1=0}^{1} \Delta f_{a_i}^{b_i} \frac{x^{i_1}(1-x)^{1-i_1}}{B(i_1+1, 2-i_1)} \frac{y^{i_2}(1-y)^{2-i_2}}{B(i_2+1, 3-i_2)}. \tag{9}$$

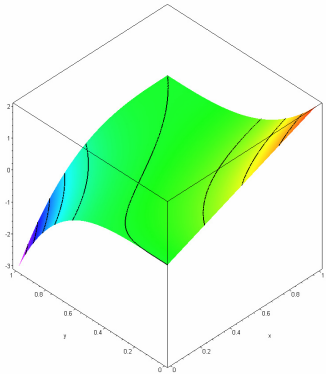 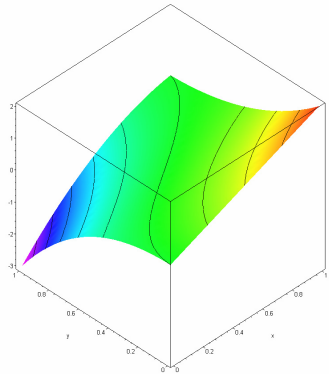 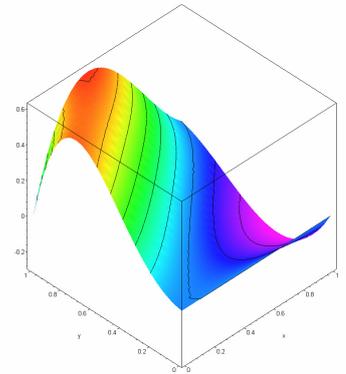

Fig.1   Fig. 2   Fig. 3

$f(x,y)$   $B_n f(x,y)$   $f(x,y) - B_n f(x,y)$

A direct consequence of Proposition 1 concerns the monotonicity behaviour of multivariate Bernstein polynomials.

**Definition 2.** Let $g$ be a real-valued function on $\mathbb{R}^d$. We call $g$ $d$-monotone iff $\Delta g_{\mathbf{a}}^{\mathbf{b}} \geq 0$ for all $\mathbf{a} = (a_1, \cdots, a_d)$, $\mathbf{b} = (b_1, \cdots, b_d) \in \mathbb{R}^d$ with $a_i < b_i$, $1 \leq i \leq d$.

It is obvious by the iterated mean value theorem that for a sufficiently smooth function $g$, $d$-monotonicity is equivalent to

$$\frac{\partial^d}{\partial x_1 \cdots \partial x_d} g(x_1, \cdots, x_d) \geq 0 \text{ for all } (x_1, \cdots, x_d) \in \mathbb{R}^d. \tag{10}$$

Note that in case that $g$ is a $d$-dimensional cumulative distribution function of a probability measure $P$ on the $d$-dimensional Borel $\sigma$-field $\mathcal{B}^d$, then $\Delta g_{\mathbf{a}}^{\mathbf{b}} = P\left[\underset{i=1}{\overset{d}{\times}}(a_i, b_i]\right]$.

**Proposition 2.** Let $f$ be a real-valued $d$-monotone function on $\mathbb{R}^d$. Then the corresponding multivariate Bernstein polynomial $B_{\mathbf{n}} f$ is also $d$-monotone. In particular, the Bernstein polynomial density $b_{\mathbf{n}} f$ is a positive-linear combination of product beta densities.

**Proof.** By the arguments above and the notation as in Proposition 1, we have

$$b_{\mathbf{n}} f(x_1, \cdots, x_d) := \frac{\partial^d}{\partial x_1 \cdots \partial x_d} B_{\mathbf{n}} f(x_1, \cdots, x_d) = \sum_{i_d=0}^{n_d-1} \cdots \sum_{i_1=0}^{n_1-1} \Delta f_{\mathbf{a}_i}^{\mathbf{b}_i} \prod_{j=1}^{d} f_{\text{beta}}(x_j; i_j+1, n_j - i_j) \geq 0 \tag{11}$$

which is a sufficient condition for $B_{\mathbf{n}} f$ to be $d$-monotone. ●

Note that the polynomial $f$ from Example 1 is not 2-monotone since $\Delta f_{\mathbf{a}}^{\mathbf{b}} = -.00126... < 0$ with $\mathbf{a} = (0.2, 0.4)$ and $\mathbf{b} = (0.27, 0.45)$. However, the slightly modified polynomial $g(x, y) = f(x, y) + 6xy$ is 2-monotone since $\frac{\partial^2}{\partial x \partial y} g(x, y) = 6 - 6(1-y)^2 + 36(1-x)^2 y^3 \geq 0$ with the unique global minimum point $(x, y) = (1, 1)$ and $\frac{\partial^2}{\partial x \partial y} g(1, 1) = 0$. With respect to the corresponding multivariate Bernstein polynomial, we now have which also explicitly shows that the Bernstein polynomial $B_{\mathbf{n}} g$ is 2-monotone.

| $i_1$ | 0 | 1 | 0 | 1 | 0 | 1 |
|---|---|---|---|---|---|---|
| $i_2$ | 0 | 0 | 1 | 1 | 2 | 2 |
| $\Delta g_{\mathbf{a}_i}^{\mathbf{b}_i}$ | $\dfrac{71}{216}$ | $\dfrac{65}{216}$ | $\dfrac{265}{216}$ | $\dfrac{175}{216}$ | $\dfrac{221}{72}$ | $\dfrac{91}{72}$ |

Tab. 2

## 3. From Bernstein polynomials to Bernstein copulas

**Remark 1.** Seemingly Proposition 2 can be usefully applied to arbitrary $d$-dimensional cumulative distribution functions $F$ concentrated on the unit cube $\mathbf{C}_d := [0,1]^d$ (continuous or not) such that the corresponding multivariate Bernstein polynomial

$$B_{\mathbf{n}}F(x_1,\cdots,x_d) = \sum_{i_d=0}^{n_d}\cdots\sum_{i_1=0}^{n_1} F\left(\frac{i_1}{n_1},\cdots,\frac{i_d}{n_d}\right)\prod_{j=1}^{d}\binom{n_j}{i_j}x_j^{i_j}(1-x_j)^{n_j-i_j}, \quad \mathbf{x}=(x_1,\cdots,x_d)\in\mathbf{C}_d \quad (12)$$

also is a cumulative distribution function since $B_{\mathbf{n}}F$ is non-negative and $d$-increasing with $B_{\mathbf{n}}F(0,\cdots,0) = F(0,\cdots,0)$ and $B_{\mathbf{n}}F(1,\cdots,1) = F(1,\cdots,1) = 1$. In particular, the Bernstein polynomial density $b_{\mathbf{n}}F$ always is a (probabilistic) mixture of product beta densities as explicitly noted in [6] and [23] for Bernstein copulas. Note also that this observation was the motivation for the setup in [20].

**Example 2.** We consider a two-dimensional random vector $\mathbf{X}=(X,Y)$ with a discrete distribution concentrated on $\{\mathbf{x},\mathbf{y}\}$ with $\mathbf{x}=(x_1,x_2)=(0.2,0.7)$ and $\mathbf{y}=(y_1,y_2)=(0.3,0.5)$ given by

| $P(X=x_i, Y=y_i)$ |     | $x_1$ | $x_2$ |
|---|---|---|---|
|   |     | 0.2 | 0.7 |
| $y_1$ | 0.3 | 0.3 | 0.2 |
| $y_2$ | 0.5 | 0.2 | 0.3 |

Tab. 3

The following graphs show the corresponding cumulative distribution function $F$ as well as the corresponding Bernstein polynomials $B_{\mathbf{n}}F$ and densities $b_{\mathbf{n}}F$ for various choices of $\mathbf{n}$.

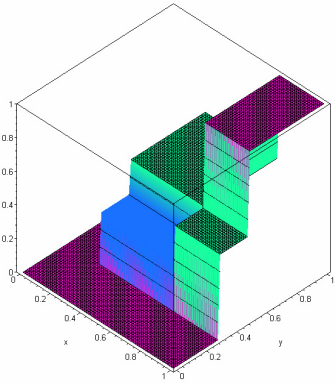 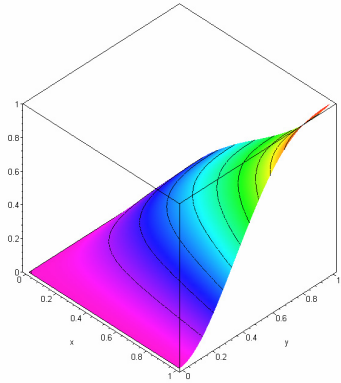 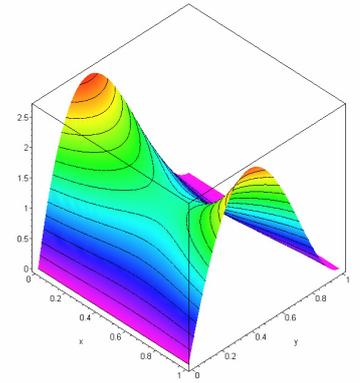

Fig. 4          Fig. 5          Fig. 6

$F(x,y)$        $B_{\mathbf{n}}F(x,y)$        $b_{\mathbf{n}}F(x,y)$

$n_1 = 3, \; n_2 = 5$

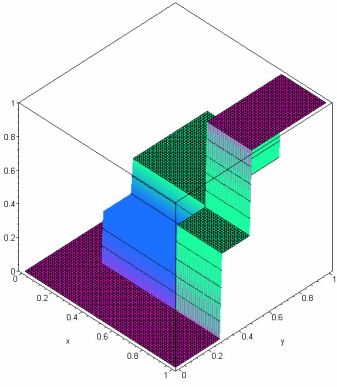 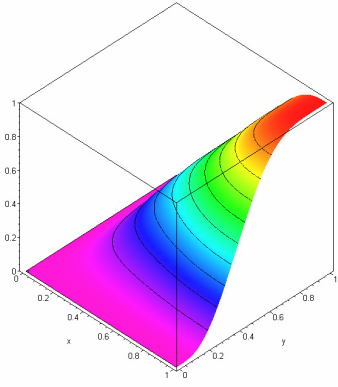 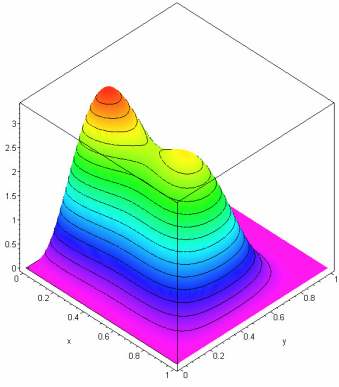

| Fig. 7 | Fig. 8 | Fig. 9 |
| --- | --- | --- |
| $F(x,y)$ | $B_{\mathbf{n}}F(x,y)$ | $b_{\mathbf{n}}F(x,y)$ |

$$n_1 = 11, \ n_2 = 7$$

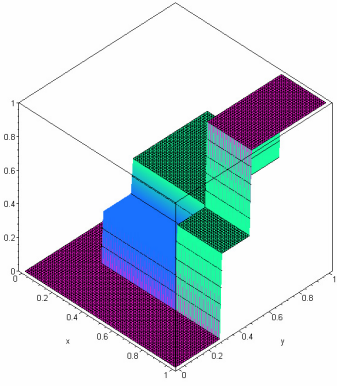 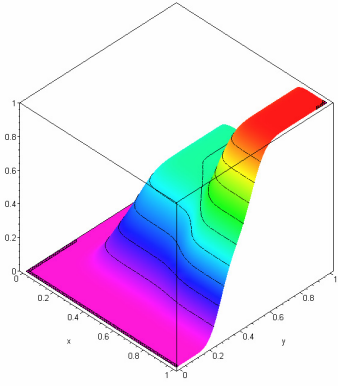 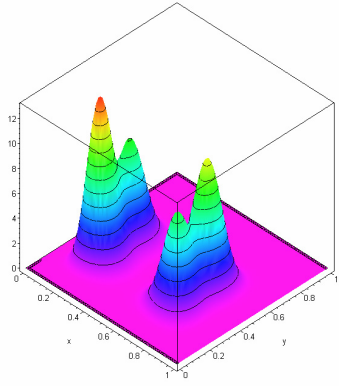

| Fig. 10 | Fig. 11 | Fig. 12 |
| --- | --- | --- |
| $F(x,y)$ | $B_{\mathbf{n}}F(x,y)$ | $b_{\mathbf{n}}F(x,y)$ |

$$n_1 = 50, \ n_2 = 50$$

The above figures clearly visualize the approximation effect of multivariate Bernstein polynomials for discrete multivariate distributions if $\min(n_1,\cdots,n_d)$ gets large. In particular, the Bernstein polynomial density has spikes around the support points of the underlying discrete distribution.

To simplify notation, we will use the following convention. Let $d > 1$ be a natural number and $\mathbf{x} = (x_1,\cdots,x_d) \in \mathbb{R}^d$ be arbitrary. Then, for $y \in \mathbb{R}$, let

$$\mathbf{x}_{\to k}(y) := \begin{cases} (y, x_2, \cdots, x_d) & \text{if } k = 1 \\ (x_1, \cdots, x_{k-1}, y, x_{k+1}, \cdots, x_d) & \text{if } 1 < k < d \\ (x_1, \cdots, x_{d-1}, y) & \text{if } k = d \end{cases} \qquad (13)$$

denote the vector $\mathbf{x}$ where the $k$-th component is replaced by $y$.

**Proposition 3.** Suppose that for $d > 1$ there is a cumulative distribution function $F:[0,1]^d \to [0,1]$ with $F(0,\cdots,0) = 0$ and $F(1,\cdots,1) = 1$ such that for given natural numbers $n_1,\cdots,n_d > 1$ we have $F\left(\mathbf{1}_{\to k}\left(\dfrac{i}{n_j}\right)\right) = \dfrac{i}{n_j}$ for $i \in \{0,1,\cdots,n_j\}$, $j = 1,\cdots,d$, $k = 1,\cdots,d$, where $\mathbf{1} = (1,\cdots,1) \in \mathbb{R}^d$. Then the $d$-dimensional Bernstein polynomial $B_\mathbf{n} F$ with $\mathbf{n} = (n_1,\cdots,n_d)$ associated with $F$ is a copula.

**Proof:** By Remark 1 above we know that $B_\mathbf{n} F$ also is a cumulative distribution function with $B_\mathbf{n} F(0,\cdots,0) = F(0,\cdots,0) = 0$ and $B_\mathbf{n} F(1,\cdots,1) = F(1,\cdots,1) = 1$, and (note that $0^0 = 1$)

$$B_\mathbf{n} F(\mathbf{1}_{\to k} x) = \sum_{i_d=0}^{n_d}\cdots\sum_{i_1=0}^{n_1} F\left(\frac{i_1}{n_1},\cdots,\frac{i_d}{n_d}\right) \prod_{j=1}^{d}\binom{n_j}{i_j} x_j^{i_j}(1-x_j)^{n_j-i_j}$$

$$= \sum_{i=0}^{n_k}\binom{n_k}{i} F\left(\mathbf{1}_{\to k}\left(\frac{i}{n_k}\right)\right) x^i (1-x)^{n_k-i} = \frac{1}{n_k}\sum_{i=0}^{n_k} i\binom{n_k}{i} x^i (1-x)^{n_k-i} = \frac{n_k x}{n_k} = x \quad (14)$$

for $k = 1,\cdots,d$ and $0 \leq x \leq 1$ ($n_k x$ is the expectation of the Binomial distribution with $n_k$ trials and success probability $x$). The marginal distributions induced by $B$ hence are continuous uniform, which means that $B$ is indeed a copula. ●

Note that Proposition 3 was already implicitly formulated in [6] and [17], see also [8], Chapter 4.1.2. We reformulate the corresponding statements there in an appropriate way.

**Corollary 1.** Let $\mathbf{U} = (U_1,\cdots,U_d)$ be a discrete random vector whose marginal component $U_i$ follows a discrete uniform distribution over $T_i := \{0,1,\cdots,n_i - 1\}$ with integers $n_i > 1$, $i = 1,\cdots,d$. Then the multivariate Bernstein polynomial $B_\mathbf{n} F$ derived from the cumulative distribution function $F$ for the scaled random vector $\mathbf{V} = \left(\dfrac{U_1+1}{n_1},\cdots,\dfrac{U_d+1}{n_d}\right)$ given by $F(x_1,\cdots,x_d) = P(V_1 \leq x_1,\cdots,V_d \leq x_d)$, $\mathbf{x} = (x_1,\cdots,x_d) \in \mathbf{C}_d$ is a copula. The corresponding Bernstein copula density $b_\mathbf{n} F$ is given by

$$b_\mathbf{n} F(x_1,\cdots,x_d) = \sum_{i_d=0}^{n_d-1}\cdots\sum_{i_1=0}^{n_1-1} P(\mathbf{U} = (i_1,\cdots,i_d)) \prod_{j=1}^{d} f_{\text{beta}}(x_j; i_j+1, n_j-i_j),\ (x_1,\cdots,x_d) \in \mathbf{C}_d. \quad (15)$$

**Proof.** For $i_j \in T_j$, $j = 1,\cdots,d$ we have $F\left(\dfrac{i_1}{n_1},\cdots,\dfrac{i_d}{n_d}\right) = P\left(V_1 \leq \dfrac{i_1}{n_1},\cdots,V_d \leq \dfrac{i_d}{n_d}\right) = P(U_1 < i_1,\cdots,U_d < i_d)$

and hence $\Delta F_{\mathbf{a}_i}^{\mathbf{b}_i} = P\left(\dfrac{i_1}{n_1} < V_1 \leq \dfrac{i_1+1}{n_1},\cdots,\dfrac{i_d}{n_d} < V_d \leq \dfrac{i_d+1}{n_d}\right) = P(\mathbf{U} = (i_1,\cdots,i_d))$. ●

**Remark 2.** We call $B_\mathbf{n} F$ the Bernstein copula induced by $\mathbf{U}$. In coincidence with [17] we also call $\mathbf{U}$ the discrete skeleton of the Bernstein copula $B_\mathbf{n} F$ and the number $n_1 \times \cdots \times n_d$ its grid size. If $\mathbf{V}$ is an arbitrary discrete random vector over $\mathbf{T} := \underset{i=1}{\overset{d}{\times}} T_i$, we call $\mathbf{V}$ an admissible discrete skeleton if the marginal distributions are discrete uniform. So every admissible skeleton over $\mathbf{T}$ induces a corresponding Bernstein copula via the multivariate Bernstein polynomial of its rescaled cumulative distribution function. The corresponding Bernstein copula density is a mixture of product beta kernels with weights given by the individual probabilities representing the admissible skeleton.

## 4. Empirical Bernstein copulas

Bernstein copulas can be easily constructed from independent samples $\mathbf{X}_1,\cdots,\mathbf{X}_n$, $n \in \mathbb{N}$ of $d$-dimensional random vectors with the same intrinsic dependence structure and the same marginal distributions. For simplicity, we assume here that the marginal distributions are continuous in order to avoid ties in the observations. The simplest way to construct an empirical Bernstein copula is on the basis of Deheuvel's empirical copula [7] in the form of a cumulative distribution function which can be represented by an admissible discrete skeleton derived from the individual ranks $r_{ij}$, $i=1\cdots,d$, $j=1,\cdots,n$ of the observation vectors $\mathbf{x}_j = (x_{1j},\cdots,x_{dj})$, $j=1,\cdots,n$, given by the order statistics $x_{i,r_{i1}} < x_{i,r_{i2}} < \cdots < x_{i,r_{in}}$, i.e. $r_{ij} = k$ iff $x_{ij}$ is the $k$-largest value of the $i$-th observed component. The discrete skeleton $\mathbf{U}$ is here given by a random vector over $\mathbf{T} := \{0,1,\cdots,n-1\}^d$ with a discrete uniform distribution over the $n$ support points $\mathbf{s}_1,\cdots,\mathbf{s}_n$ where $\mathbf{s}_j = (r_{1j}-1,\cdots,r_{dj}-1)$, $j=1,\cdots,n$. Since the empirical copula converges in distribution to the true underlying copula when $n \to \infty$ it follows that the corresponding empirical Bernstein copula does so likewise, cf. [8], Chapter 4.1.2. This provides – in the light of relation (15) – in particular a simple way of generating samples from an empirical Bernstein copula by Monte Carlo methods in two steps:

Step 1: Select an index $N$ randomly and uniformly among $1,\cdots,n$.

Step 2: Generate $d$ independent beta distributed random variables $V_1,\cdots,V_d$ (also independent of $N$) where $V_i$ follows a beta distribution with parameters $r_{iN}$ and $n+1-r_{iN}$, $i=1,\cdots,d$.

Then $\mathbf{V} := (V_1,\cdots,V_d)$ is a sample point from the empirical Bernstein copula.

This has also been observed in [23], but was known earlier, see e.g. [6]. In what follows we will discuss the data set presented in [17], Section 3 in more detail.

**Example 3.** The following table contains the ranks for observed insurance data from windstorm ($i=1$) and flooding ($i=2$) losses in central Europe for 34 consecutive years discussed in [17].

| $r_{ij}$ | $j$ | 1 | 2 | 3 | 4 | 5 | 6 | 7 | 8 | 9 | 10 | 11 | 12 | 13 | 14 | 15 | 16 | 17 |
|---|---|---|---|---|---|---|---|---|---|---|---|---|---|---|---|---|---|---|
| $i$ | 1 | 1 | 2 | 3 | 4 | 5 | 6 | 7 | 8 | 9 | 10 | 11 | 12 | 13 | 14 | 15 | 16 | 17 |
|  | 2 | 12 | 5 | 31 | 7 | 24 | 18 | 17 | 3 | 2 | 19 | 10 | 9 | 21 | 15 | 14 | 4 | 6 |

| $r_{ij}$ | $j$ | 18 | 19 | 20 | 21 | 22 | 23 | 24 | 25 | 26 | 27 | 28 | 28 | 30 | 31 | 32 | 33 | 34 |
|---|---|---|---|---|---|---|---|---|---|---|---|---|---|---|---|---|---|---|
| $i$ | 1 | 18 | 19 | 20 | 21 | 22 | 23 | 24 | 25 | 26 | 27 | 28 | 29 | 30 | 31 | 32 | 33 | 34 |
|  | 2 | 34 | 1 | 23 | 11 | 29 | 33 | 13 | 8 | 20 | 32 | 28 | 22 | 16 | 26 | 25 | 30 | 27 |

Tab. 4

The following graphs show some plots for the empirical Bernstein copula.

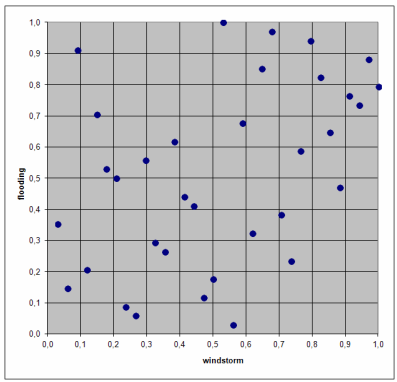 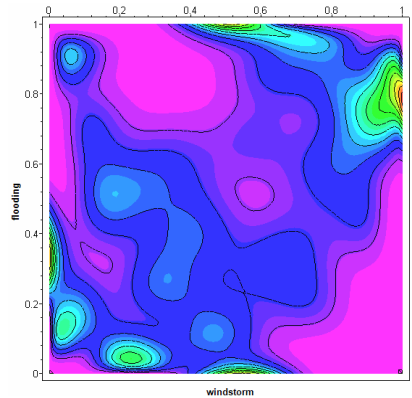 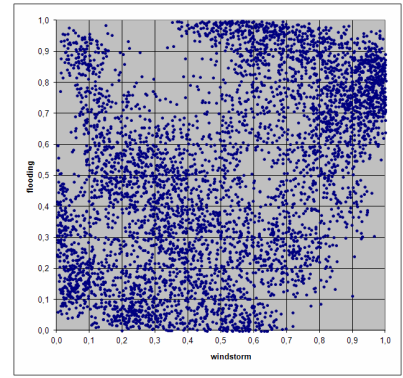

Fig. 13 | Fig. 14 | Fig. 15

support points of scaled skeleton | copula density contour plot | simulation of 5.000 copula pairs

It is clearly to be seen that the empirical Bernstein copula density is quite bumpy here, e.g. in comparison with the Gaussian copula density fitted to the data set above.

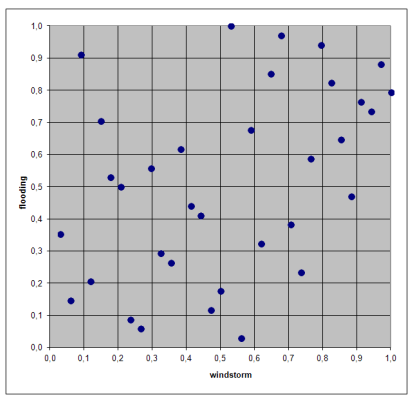 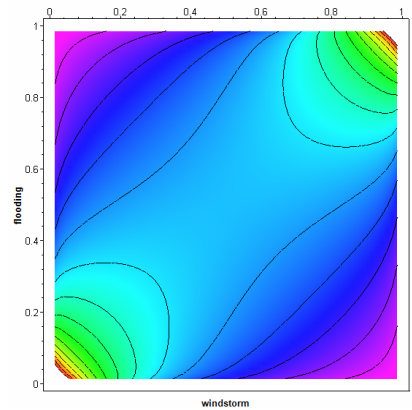 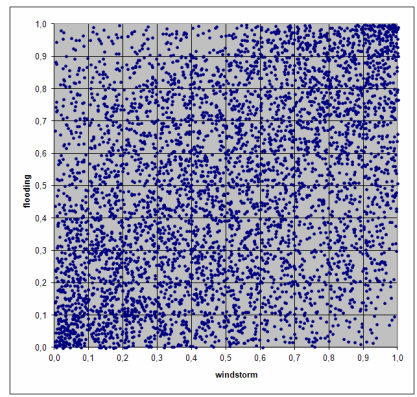

Fig. 16 | Fig. 17 | Fig. 18

support points of scaled skeleton | Gaussian copula density contour plot | simulation of 5.000 Gaussian copula pairs

From a practical point of view, it might therefore be desirable to adapt the empirical Bernstein copula to a smaller support set $\mathbf{T}^* := \bigtimes_{i=1}^{d} T_i^* \subset \mathbf{T} := \bigtimes_{i=1}^{d} T_i$ for the underlying discrete skeleton. This was the central idea in [17]. The disadvantage of the method proposed there is, however, that the number of support points of $\mathbf{U}^*$ gets dramatically larger and is typically of exponential order with increasing sample size. We therefore propose a simpler way how to find a smaller discrete approximating skeleton $\mathbf{U}^*$ with an arbitrary given grid size in the subsequent chapter.

## 5. Adaptive Bernstein copula estimation

We start with the individual ranks $r_{ij}$ of the observation vectors $\mathbf{x}_j = (x_{1j}, \cdots, x_{dj})$, $j = 1, \cdots, n$. Let $\mathbf{U}$ denote the canonical admissible discrete skeleton as described in the preceding chapter, derived from the empirical copula. Our aim is to find a good approximating admissible discrete skeleton $\mathbf{U}^*$ with a given grid size $n_1 \times \cdots \times n_d$ where the $n_i$ are typically smaller than $n$. We proceed in two steps:

### 1. Step: Augmentation

Select an integer $M$ such that all $n_i$, $i = 1, \cdots, d$ are divisors of $M$, for instance their least common multiple. We construct pseudo-ranks $r_{ij}^+$ in the following way:

$$r_{ij}^+ := r_{i, \lceil \frac{j}{M} \rceil} \cdot M + \left( \left\lceil \frac{j}{M} \right\rceil - 1 \right) \cdot M + 1 - j, \quad i = 1, \cdots, d, \ j = 1, \cdots, M \cdot n. \tag{16}$$

Here $\lceil x \rceil := \min\{m \in \mathbb{Z} \mid x \leq m\}$, $x \in \mathbb{R}$ stands for "rounding up". Let $\mathbf{U}^+ = (U_1^+, \cdots, U_d^+)$ be the uniformly discretely distributed random vector over $\{0, 1, \cdots, M \cdot n - 1\}^d$ with support points $\mathbf{s}_1, \cdots, \mathbf{s}_{M \cdot n}$ where $\mathbf{s}_j = (r_{1j}^+ - 1, \cdots, r_{dj}^+ - 1)$, $j = 1, \cdots, n \cdot M$. Note that the probability mass is $\frac{1}{M \cdot n}$ for each support point, and that $\mathbf{U}^+$ is an admissible discrete skeleton.

### 2. Step: Reduction

Construct the final ranks $r_{ij}^*$ in the following way:

$$r_{ij}^* := \left\lceil \frac{r_{ij}^+ \cdot n_i}{n \cdot M} \right\rceil, \quad i = 1, \cdots, d, \ j = 1, \cdots, M \cdot n. \tag{17}$$

It follows from the above definition, that there will be replicates in the final ranks and that $r_{ij}^* - 1$ takes values in the set $T_i^* = \{0, 1, \cdots, n_i - 1\}$. A point $\mathbf{s} = (s_1, \cdots, s_d)$ will be a support point of the final admissible skeleton $\mathbf{U}^*$ if there exist final ranks such that $\mathbf{s} = (r_{1, j_1}, \cdots, r_{d, j_d})$ for some $j_1, \cdots, j_d \in \{1, \cdots, M \cdot n\}$. The probability mass attached to $\mathbf{s}$ is given by the number $\frac{K}{M \cdot n}$ where $K$ is the number of rank combinations $(r_{1, j_1}, \cdots, r_{d, j_d})$ that lead to the same $\mathbf{s}$. This also enables very simple Monte Carlo realisations of the corresponding Bernstein copula as described in chapter 4 by first selecting an index $N$ randomly and uniformly among $1, \cdots, M \cdot n$ and then by proceeding as in step 2 there with all of the $r_{ij}^*$.

Note that the above augmentation step creates permutations of the set $\{1, \cdots, M \cdot n\}$ in each component so that the pseudo-ranks $r_{ij}^+$ actually lead to an admissible discrete skeleton, cf. [6], chapter 4.
The mathematical correctness of the reduction step follows from the proof of Proposition 2.5 in [23].

In the augmentation step, $M$-wise partial permutations would not change the result but would create a more "random" augmentation of the original ranks.

**Example 4.** Consider the following rank table with $n = 5$:

| $r_{ij}$ | | i | | probabilities $p(r_1, r_2)$ |
|---|---|---|---|---|
| | | 1 | 2 | |
| j | 1 | 1 | 3 | 0.2 |
| | 2 | 2 | 4 | 0.2 |
| | 3 | 3 | 1 | 0.2 |
| | 4 | 4 | 2 | 0.2 |
| | 5 | 5 | 5 | 0.2 |

Tab. 5

We want to create approximate final ranks with $n_1 = 3$ and $n_2 = 4$. Both numbers are not a divisor of $n$, so we choose $M = 3 \cdot 4 = 12$. We show a part of the resulting pseudo-ranks:

| $r_{ij}^+$ | | i | | probabilities $p(r_1^+, r_2^+)$ |
|---|---|---|---|---|
| | | 1 | 2 | |
| j | 1 | 12 | 36 | $0.01\overline{6}$ |
| | 2 | 11 | 35 | $0.01\overline{6}$ |
| | 3 | 10 | 34 | $0.01\overline{6}$ |
| | ⋮ | ⋮ | ⋮ | ⋮ |
| | 13 | 24 | 48 | $0.01\overline{6}$ |
| | 14 | 23 | 47 | $0.01\overline{6}$ |
| | 15 | 22 | 46 | $0.01\overline{6}$ |
| | ⋮ | ⋮ | ⋮ | ⋮ |
| | 25 | 36 | 12 | $0.01\overline{6}$ |
| | 26 | 35 | 11 | $0.01\overline{6}$ |
| | 27 | 34 | 10 | $0.01\overline{6}$ |
| | ⋮ | ⋮ | ⋮ | ⋮ |
| | 58 | 51 | 51 | $0.01\overline{6}$ |
| | 59 | 50 | 50 | $0.01\overline{6}$ |
| | 60 | 49 | 49 | $0.01\overline{6}$ |

Tab. 6

For the final ranks we obtain the following table:

| $r_{ij}^*$ | | $i$ | | |
|---|---|---|---|---|
| | | 1 | 2 | probabilities $p(r_1^*, r_2^*)$ |
| $j$ | 1 | 1 | 2 | 0.1 |
| | 2 | 1 | 3 | $0.2\bar{3}$ |
| | 3 | 2 | 1 | 0.25 |
| | 4 | 2 | 2 | $0.01\bar{6}$ |
| | 5 | 2 | 3 | $0.01\bar{6}$ |
| | 6 | 2 | 4 | 0.05 |
| | 7 | 3 | 2 | $0.1\bar{3}$ |
| | 8 | 3 | 4 | 0.2 |

Tab. 7

From $\sum_{j=1}^{4} p(i,j) = 0.\bar{3},\ i=1,2,3$ and $\sum_{i=1}^{3} p(i,j) = 0.25,\ j=1,2,3,4$ we see that the induced skeleton is indeed admissible.

The following graphs show the corresponding copula densities $c_{\mathbf{U}}(x_1, x_2)$ and $c_{\mathbf{U}^*}(x_1, x_2)$ induced by $\mathbf{U}$ and $\mathbf{U}^*$.

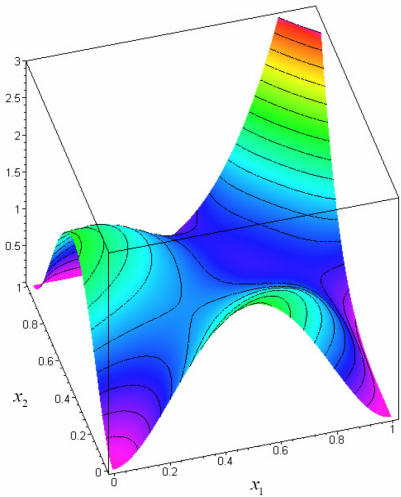
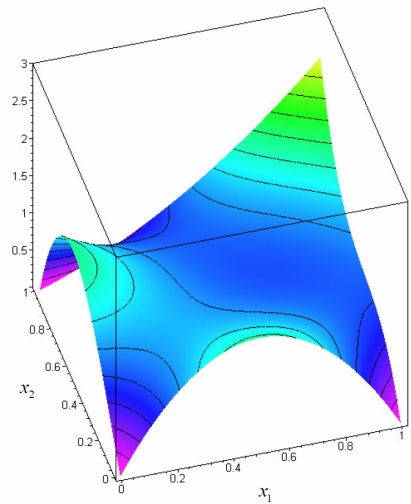

Fig. 19  
$c_{\mathbf{U}}(x_1, x_2)$

Fig. 20  
$c_{\mathbf{U}^*}(x_1, x_2)$

Seemingly the shape of both densities is similar, reflecting the structure of the original ranks quite well. However, the density $c_{\mathbf{U}^*}$ is less wiggly than the density $c_{\mathbf{U}}$, as was intended.

Note also that a reduction of complexity for copulas in the sense discussed here is also an essential topic in [12], chapter 3; see in particular Fig. 2 there. However, the underlying problem of a consistent reduction of complexity is not really discussed there.

# 6. Applications to risk management

In this chapter we want to investigate the data set of Example 3 in more detail. It was the basis data set in [17]. In particular, we want to discuss the effect of different adaptive Bernstein copula estimations on the estimation of risk measures like Value at Risk which is the basis for Solvency II, for instance.

In [17], the number $n = 34$ of the original observations was first reduced to $n_1 = n_2 = 10$ by a least squares technique. The resulting optimal discrete skeleton with probabilities $p_{ij}$, $i \in T_1^*$, $j \in T_2^*$ is presented here with $T_1^* = T_2^* = \{0, 1, \cdots, 9\}$..

| $p_{ij}$ | | | | | | $i$ | | | | | |
|---|---|---|---|---|---|---|---|---|---|---|---|
| | | 0 | 1 | 2 | 3 | 4 | 5 | 6 | 7 | 8 | 9 |
| $j$ | 9 | 0.0032 | 0.0000 | 0.0022 | 0.0000 | 0.0032 | 0.0266 | 0.0320 | 0.0274 | 0.0028 | 0.0028 |
| | 8 | 0.0318 | 0.0000 | 0.0014 | 0.0000 | 0.0024 | 0.0000 | 0.0312 | 0.0000 | 0.0020 | 0.0314 |
| | 7 | 0.0000 | 0.0000 | 0.0000 | 0.0000 | 0.0000 | 0.0000 | 0.0000 | 0.0204 | 0.0251 | 0.0545 |
| | 6 | 0.0032 | 0.0275 | 0.0022 | 0.0000 | 0.0032 | 0.0265 | 0.0026 | 0.0000 | 0.0322 | 0.0028 |
| | 5 | 0.0003 | 0.0246 | 0.0287 | 0.0215 | 0.0003 | 0.0000 | 0.0000 | 0.0246 | 0.0000 | 0.0000 |
| | 4 | 0.0034 | 0.0278 | 0.0024 | 0.0246 | 0.0034 | 0.0000 | 0.0029 | 0.0000 | 0.0324 | 0.0030 |
| | 3 | 0.0266 | 0.0000 | 0.0000 | 0.0000 | 0.0266 | 0.0206 | 0.0261 | 0.0000 | 0.0000 | 0.0000 |
| | 2 | 0.0034 | 0.0000 | 0.0025 | 0.0540 | 0.0034 | 0.0000 | 0.0029 | 0.0277 | 0.0031 | 0.0031 |
| | 1 | 0.0252 | 0.0201 | 0.0000 | 0.0000 | 0.0546 | 0.0000 | 0.0000 | 0.0000 | 0.0000 | 0.0000 |
| | 0 | 0.0029 | 0.0000 | 0.0607 | 0.0000 | 0.0029 | 0.0263 | 0.0023 | 0.0000 | 0.0025 | 0.0025 |

Tab. 8

An application of the adaptive strategy described in the preceding chapter gives alternatively the following less complex table. Here we have chosen $M = 5$.

| $p_{ij}^*$ | | | | | | $i$ | | | | | |
|---|---|---|---|---|---|---|---|---|---|---|---|
| | | 0 | 1 | 2 | 3 | 4 | 5 | 6 | 7 | 8 | 9 |
| $j$ | 9 | 0.0118 | 0.0000 | 0.0000 | 0.0000 | 0.0000 | 0.0294 | 0.0294 | 0.0294 | 0.0000 | 0.0000 |
| | 8 | 0.0176 | 0.0000 | 0.0000 | 0.0000 | 0.0000 | 0.0000 | 0.0294 | 0.0000 | 0.0235 | 0.0294 |
| | 7 | 0.0000 | 0.0059 | 0.0000 | 0.0000 | 0.0000 | 0.0000 | 0.0000 | 0.0059 | 0.0176 | 0.0706 |
| | 6 | 0.0000 | 0.0235 | 0.0000 | 0.0176 | 0.0000 | 0.0294 | 0.0000 | 0.0000 | 0.0294 | 0.0000 |
| | 5 | 0.0000 | 0.0294 | 0.0294 | 0.0118 | 0.0000 | 0.0000 | 0.0000 | 0.0294 | 0.0000 | 0.0000 |
| | 4 | 0.0000 | 0.0235 | 0.0059 | 0.0176 | 0.0235 | 0.0000 | 0.0000 | 0.0000 | 0.0294 | 0.0000 |
| | 3 | 0.0294 | 0.0000 | 0.0000 | 0.0000 | 0.0176 | 0.0059 | 0.0412 | 0.0059 | 0.0000 | 0.0000 |
| | 2 | 0.0000 | 0.0059 | 0.0059 | 0.0529 | 0.0000 | 0.0059 | 0.0000 | 0.0294 | 0.0000 | 0.0000 |
| | 1 | 0.0412 | 0.0118 | 0.0000 | 0.0000 | 0.0471 | 0.0000 | 0.0000 | 0.0000 | 0.0000 | 0.0000 |
| | 0 | 0.0000 | 0.0000 | 0.0588 | 0.0000 | 0.0118 | 0.0294 | 0.0000 | 0.0000 | 0.0000 | 0.0000 |

Tab. 9

Seemingly the number of support points for the adaptive probabilities $p_{ij}^*$ are much less than before.

The following graphs show contour plots for the corresponding Bernstein copula densities. Here $c_1$ denotes the Bernstein copula density derived from Tab. 8, $c_2$ denotes the Bernstein copula density derived from Tab. 9. In the first case we have chosen $M = 5$, in the second case $M = 2$.

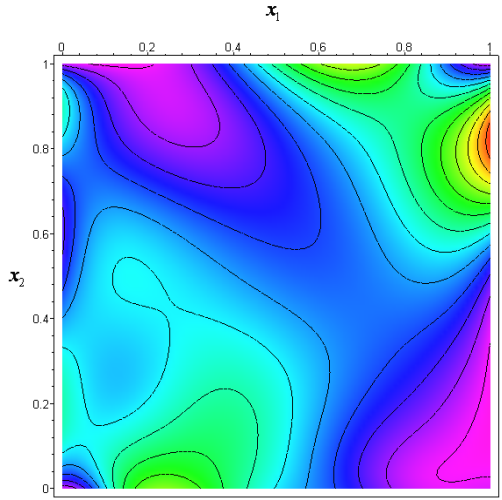
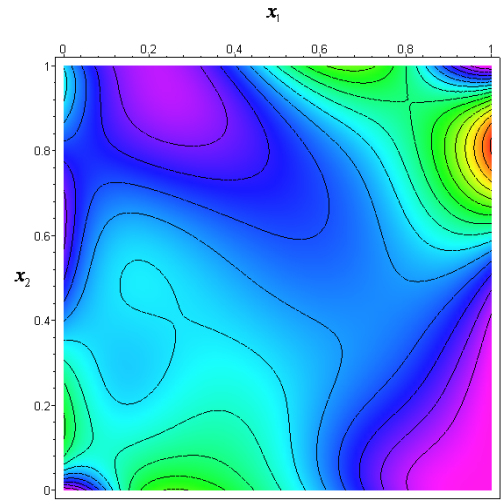

Fig. 21　　　　　　　　　　　　　　　　　　Fig. 22
$c_1(x_1, x_2)$　　　　　　　　　　　　　　　$c_2(x_1, x_2)$

Seemingly the differences are only marginal. However, in comparison with Fig. 14, the smoothing effect of the adaptive procedure is clearly visible.

The next graphs show contour plots for adaptive Bernstein copula densities $c_3$ and $c_4$ with the choices $n_1 = n_2 = 5$ and $n_1 = n_2 = 4$, respectively.

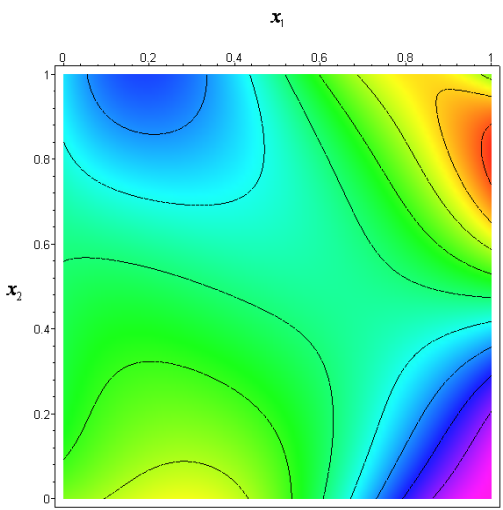
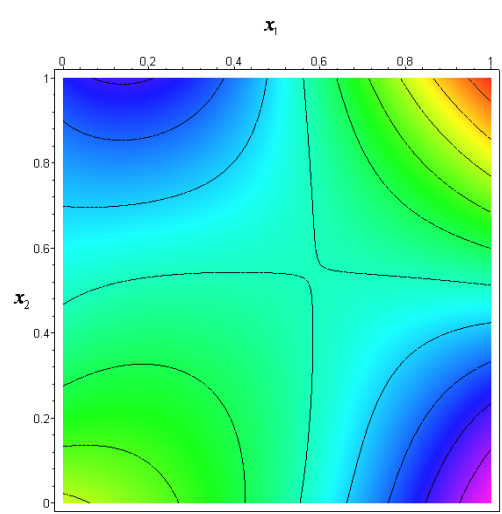

Fig. 23　　　　　　　　　　　　　　　　　　Fig. 24
$c_3(x_1, x_2)$　　　　　　　　　　　　　　　$c_4(x_1, x_2)$

In a final step, we compare estimates for the risk measure Value at Risk $VaR_\alpha$ with the risk level $\alpha = 0.5\%$ – corresponding to a return period of 200 years – on the basis of a Monte Carlo study with 1,000,000 repetitions each for the aggregated risk of windstorm and flooding losses. We consider the full Bernstein copula of Example 3 with $n_1 = n_2 = 34$ as well as the adaptive Bernstein copulas with $n_1 = n_2 = 10$, $n_1 = n_2 = 5$ and $n_1 = n_2 = 4$. For the sake of completeness, we also add estimates from the Gaussian copula, the independence as well as the co- and countermonotonicity copulas (see [8], p.11 for definitions).

The following graphs show the support points of the underlying adaptive scaled discrete skeletons as well as 5,000 simulated pairs of the adapted Bernstein copulas.

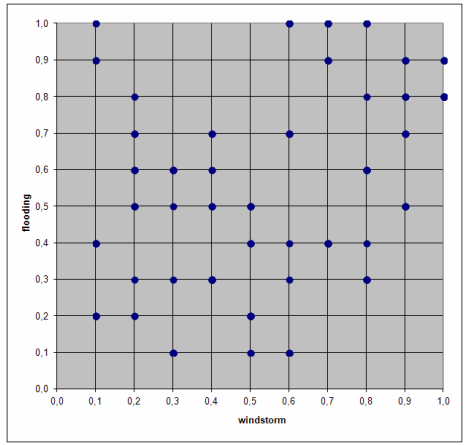
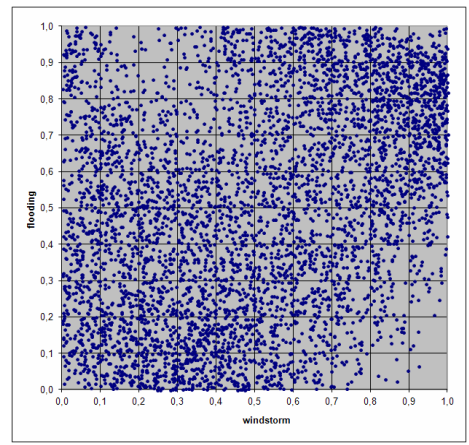

Fig. 25            Fig. 26

$n_1 = n_2 = 10$

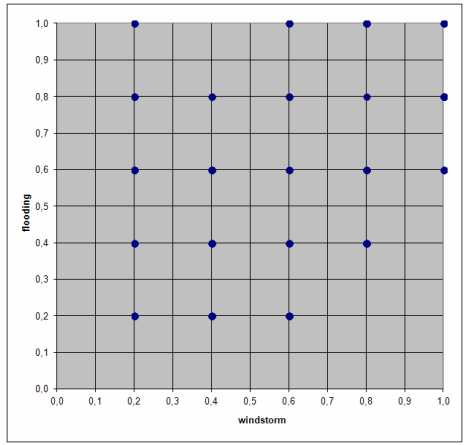
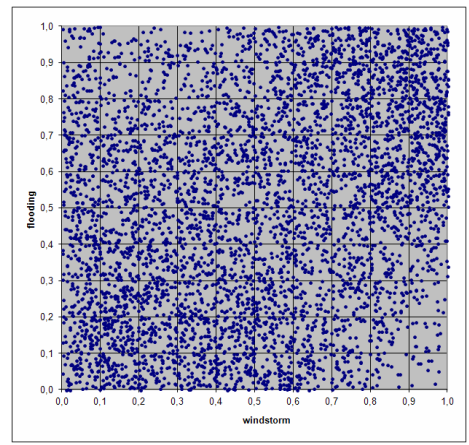

Fig. 27            Fig. 28

$n_1 = n_2 = 5$

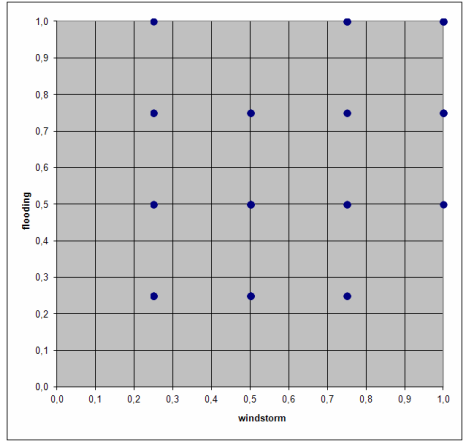
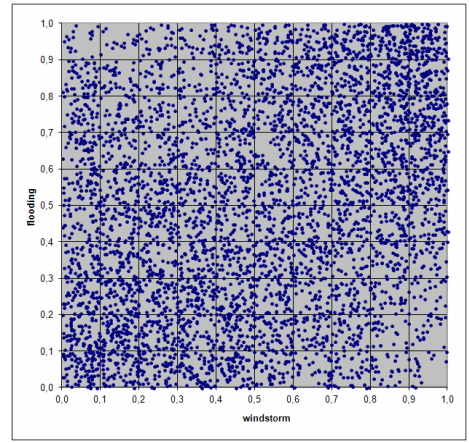

| | Fig. 29 | | Fig. 30 |

$n_1 = n_2 = 4$

The following table provides estimated values of the risk measures from the Monte Carlo simulations which are given in Mio. monetary units. For the marginal distributions, the assumptions in [17] were used.

| grid type | 34×34 | 10×10 | 5×5 | 4×4 | Gaussian | independence | comonotonic | countermonotonic |
|---|---|---|---|---|---|---|---|---|
| $VaR_{0.005}$ | 1,348 | 1,334 | 1,356 | 1,369 | 1,386 | 1,349 | 1,500 | 1,327 |

Tab. 10

Seemingly the comonotonicity copula provides the largest $VaR_{0.005}$-estimate due to an extreme tail dependence while the countermonotonicity copula provides the smallest $VaR_{0.005}$-estimate. Surprisingly, the $VaR_{0.005}$-estimates for the adaptive Bernstein copulas do not differ very much from each other and are almost identical to the estimate from the independence copula here. Note that the $VaR_{0.005}$-estimate for the Gaussian copula is slightly larger.

Significant differences are, however, visible if we look at the densities for the aggregated risk. The following graphs show empirical histograms for these densities under the models considered above, from 100,000 simulations each.

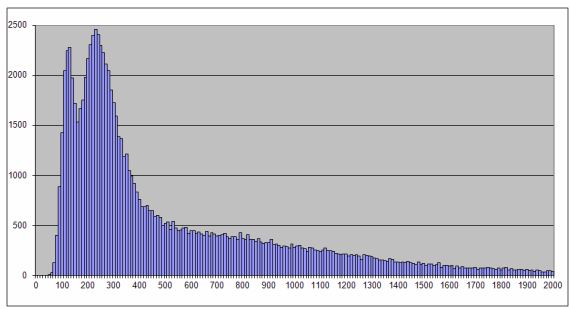
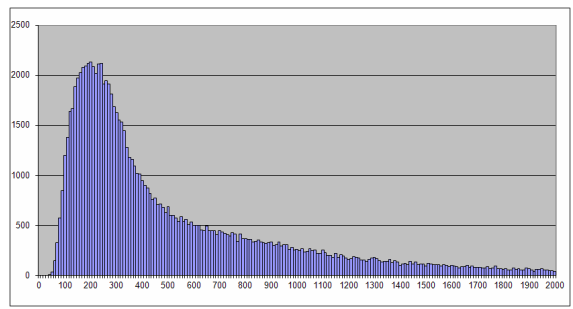

| Fig. 31 | Fig. 32 |
| Bernstein, grid type 34×34 | Bernstein, grid type 10×10 |

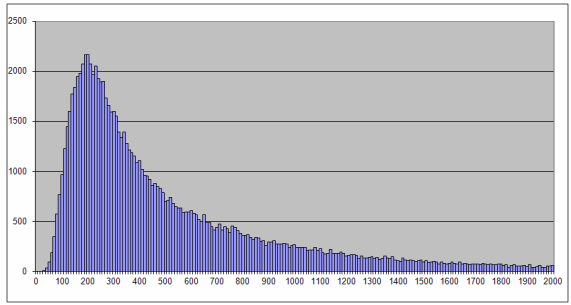

Fig. 33

Bernstein, grid type $5\times 5$

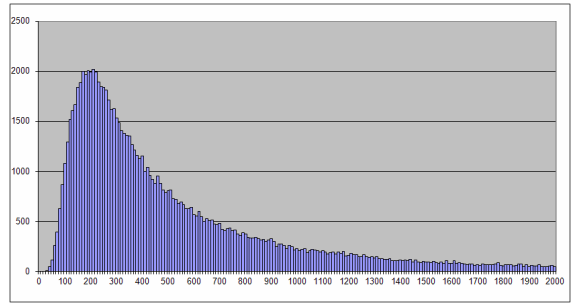

Fig. 34

Bernstein, grid type $4\times 4$

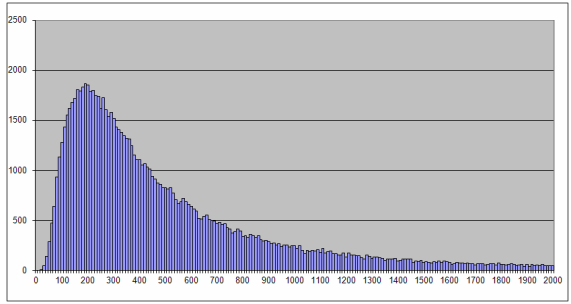

Fig. 35

Gaussian copula

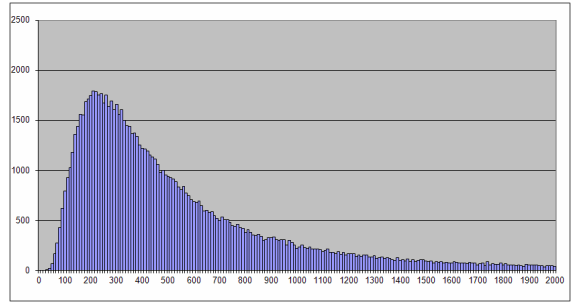

Fig. 36

independence copula

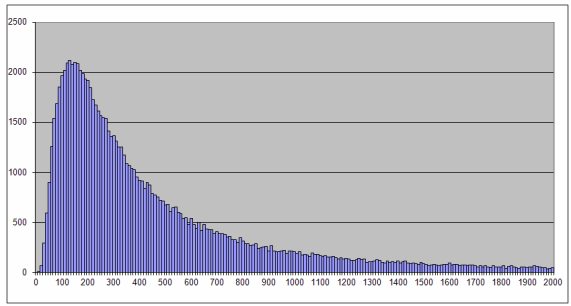

Fig. 37

comonotonicity copula

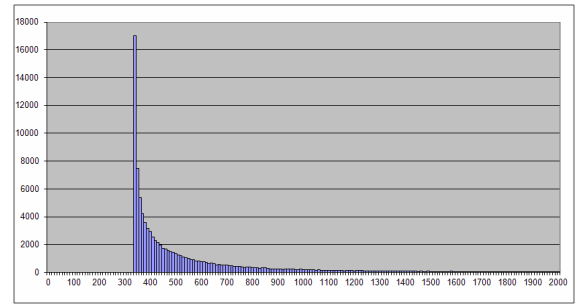

Fig. 38

countermonotonicity copula

Note that the histogram for the full Bernstein copula has two peaks, whereas the other histograms show a more smooth behaviour.

## 7. Conclusion

Adaptive Bernstein copulas are an interesting tool for smoothing the empirical dependence structure in particular in risk management applications when the number of observations is moderate to large. This prevents in particular a kind of overfitting to dependence models. They also enable Monte Carlo studies for the comparison of different estimates of risk measures or the shape of the aggregate risk distribution. If the different estimates for the risk measure do not differ much for various adaptive strategies, this could be helpful for a profound sensitivity analysis under Solvency II.

The method of reducing the complexity in the rank structures of the data could also be applied to partition-of-unity copulas, see [18], [19] and [21]. With such copulas, tail dependence can be introduced to the dependence models which cannot be obtained by Bernstein copulas due to the boundedness of the corresponding densities.

**Acknowledgement:** We thank M.Sc. Lennard Foraita for pointing out an error in a previous version of relation (16).

**Funding:** The publication of this paper was funded by the University of Oldenburg.